# Tailoring neuromorphic switching by CuN$_x$-mediated orbital currents

Tian-Yue Chen[1], Yu-Chan Hsiao[1], Wei-Bang Liao[1], and Chi-Feng Pai[1,2*]

[1]*Department of Materials Science and Engineering, National Taiwan University, Taipei 10617, Taiwan*

[2]*Center of Atomic Initiative for New Materials, National Taiwan University, Taipei 10617, Taiwan*

## Abstract

Current-induced spin-orbit torque (SOT) is regarded as a promising mechanism for driving neuromorphic behavior in spin-orbitronic devices. In principle, the strong SOT in heavy metal-based magnetic heterostructure is attributed to the spin-orbit coupling (SOC)-induced spin Hall effect (SHE) and/or the spin Rashba-Edelstein effect (SREE). Recently, SOC-free mechanisms such as the orbital angular momentum (OAM)-induced orbital Hall effect (OHE) and/or the orbital Rashba-Edelstein effect (OREE) have been proposed to generate sizable torques comparable to those from the conventional spin Hall mechanism. In this work, we show that the orbital current can be effectively generated by the nitrided light metal Cu. The overall damping-like SOT efficiency, which consists of both the spin and the orbital current contributions, can be tailored from $\xi_{\mathrm{DL}} \approx$ 0.06 to 0.4 in a Pt/Co/CuN$_x$ magnetic heterostructure by tuning the nitrogen doping concentration. Current-induced magnetization switching further verifies the efficacy of such orbital current with a critical switching current density as low as $J_\mathrm{c} \sim 5 \times 10^{10}$ A/m$^2$. Most importantly, the orbital-current-mediated memristive switching behavior can be observed in such heterostructures, which reveals that the gigantic SOT and efficient magnetization switching are the tradeoffs for the applicable window of memristive switching. Our work provides insights into the role of orbital current might play in SOT neuromorphic devices and paves a new route for making energy-efficient spin-orbitronic devices.

# I. Introduction

Current-induced spin-orbit torque (SOT) is considered as an effective mechanism to manipulate the magnetization in magnetic devices. The SOT driven magnetic random access memory (SOT MRAM) holds the advantages of fast speed, high density, and low-power consumption, which shows great potential to replace the current embedded static random access memory (SRAM) in advanced technology nodes. Conventionally, SOT is believed to be generated from the bulk spin Hall effect (SHE) [1-7] and/or the interfacial spin Rashba-Edelstein effect (SREE) [8-10] in heavy metals (HMs) and chalcogenides materials-based magnetic heterostructures with strong spin-orbit coupling (SOC), with the charge-to-spin conversion efficiencies, SOT efficiencies, ranging from 10% to over 100% [11,12]. In addition to the materials with pronounced SOC, light metals (LMs) with weak SOC are recently discovered to generate non-negligible SOTs, which attract much attention [13-26]. These "SOC-free" charge-to-spin conversions are recently recognized to originate from the orbital angular momentum (OAM)-induced orbital Hall effect (OHE) [20-23] and/or the orbital Rashba-Edelstein effect (OREE) [17-19]. The OAM dominating SOT generation has been experimentally explored in various LM-based systems such as Ti [22], Cr [20], and Cu [17-19]. This effective charge-to-spin conversion from the SOC-free OAM, in addition to its conventional SOC-based counterpart, could be a promising route to further reduce the energy consumption of spin-orbitronic devices for next generation non-volatile memories.

Besides the standard memory applications, energy-efficient spintronic and spin-orbitronic devices also show great potentials in neuromorphic computing [27-31]. To date, various types of materials/memory systems have been proposed to demonstrate neuromorphic computing, such as ferroelectric materials systems [32,33], phase change memories [34,35], and resistive memories [36,37]. Spintronic device is another prevailing option due to its versatile features such as stochasticity [38], nonlinearity [39], and nonvolatility [40]. Specifically, the memristive switching behavior in spintronic devices is found to mimic the synapse and spiking neurons by employing

SOT to manipulate domain wall propagation [28,31] and domain nucleation [27], which makes it attractive for neuromorphic applications.

To examine the feasibility of the orbital current in tuning memristive SOT switching, a series of perpendicularly magnetized Pt/Co/CuN$_x$ magnetic heterostructures are prepared. The Cu-based capping layer is chosen due to its low resistivity and the weak SOC nature of Cu makes it a suitable candidate to focus on orbital effects. Nevertheless, it has been studied that the SOT from pure Cu is too small to drive magnetization dynamics. Hence, various approaches have been adopted to strengthen the SOT from Cu. For instance, introducing dopants like Bi improves the bulk SHE by increasing the extrinsic skew scattering [41]. Naturally-oxidized Cu is also demonstrated to produce a sizable SOT from the activation of OREE [18]. Beyond oxidation, nitridation is recognized as an effective method to engineer the SOT efficiency in 5$d$ metals like Ta [42] and Pt [43]. However, the nitridation approach is yet to be performed on LMs with weak SOC like Cu. Therefore, the orbital current generation in CuN$_x$ is worth being scrutinized.

In this work, we systematically investigate the nitridation effects on the magnetic anisotropy, damping-like SOT efficiency, and neuromorphic current-induced switching behavior in Pt/Co/CuN$_x$ magnetic heterostructures. The introduction of nitrogen is found to be detrimental to the perpendicular magnetic anisotropy (PMA) in the Pt/Co/Cu(N) magnetic system. The out-of-plane (OOP) coercivity reduces rapidly as the nitrogen is introduced into Cu. The SOT characterization with hysteresis loop shift measurement then reveals that the overall damping-like SOT efficiency in Pt/Co/CuN$_x$ magnetic heterostructures can be enhanced from $\xi_{DL} \approx$ 0.06 to 0.4 by tuning the nitrogen content with a doping concentration of $Q$ = 6.3% to 28.6%. The substantial SOT reinforcement is attributed to the sizable OAM-induced orbital current generated in the CuN$_x$ layer. Current-induced switching measurement further confirms that the orbital current enhanced SOT can effectively manipulate the Co magnetization with a critical switching current density as low as $J_c \sim 5 \times 10^{10}$ A/m$^2$ for $Q$ = 28.6%. Most importantly, by varying the maximum applied current, a robust multi-states memristive SOT switching behavior mediated by the orbital current is observed. However, the memristive switching demonstration imposes an underlying challenge that the memristive window, defined as the maximum and the minimum current densities to obtain intermediate states, will be reduced as the overall SOT efficiency is increased. Our work therefore provides insights to the optimization of energy efficient SOT neuromorphic devices utilizing both spin and orbital currents.

## II. Spin current and orbital current generation

The spin current generation from conventional SOC can either originate from the bulk SHE or the interfacial SREE. The SHE describes the spin polarization separation as illustrated in Fig. 1(a) and it originates from either the intrinsic Berry curvature or the extrinsic side-jump and skew scattering [1,4]. The SREE reflects that the spin-degenerate bands would split with opposite spin polarizations in $k$-space when an electric current is applied, as shown in Fig. 1(b) [8,9]. Both the SHE and SREE follow the restriction of $J_{\text{spin}} \propto J_\text{c} \times \sigma_{\text{SH}}$, where $J_{\text{spin}}$, $J_\text{c}$, and $\sigma_{\text{SH}}$ are the spin current, charge current, and spin polarization. Their counterparts of producing orbital currents are the OHE and the OREE. The OHE comes from the orbital texture in the materials systems with multi-orbit [21]. Similar to the SHE, it reflects the orbital polarization separation, as shown in Fig 1(c). As for the OREE, analogous to the SREE, refers to an OAM-dependent energy splitting that generates a chiral orbital texture in $k$-space [17]. The OHE and OREE follow the rule of $J_{\text{orbital}} \propto J_\text{c} \times \sigma_{\text{OH}}$, where $J_{\text{orbital}}$ is the orbital current and $\sigma_{\text{OH}}$ is the orbital polarization. Despite the similarities to the SOC-dependent effects (SHE and SREE), the OAM-induced OHE and OREE are independent of SOC. In contrast to the spin current, the orbital current is forbidden to directly transfer the angular momentum to the magnetization due to the lack of exchange coupling between the OAM ($L$) and the local magnetization [20,24]. The $L$ is required to convert into the spin angular momentum ($S$) through SOC in the adjacent ferromagnet so that a torque can be exerted on the magnetic moment.

Note that this conversion process can cause discrepancies of the estimated magnitude and sign of the so-called orbital torque, which is correlated to the *L-S* conversion coefficient ($\eta$) [16,20,24].

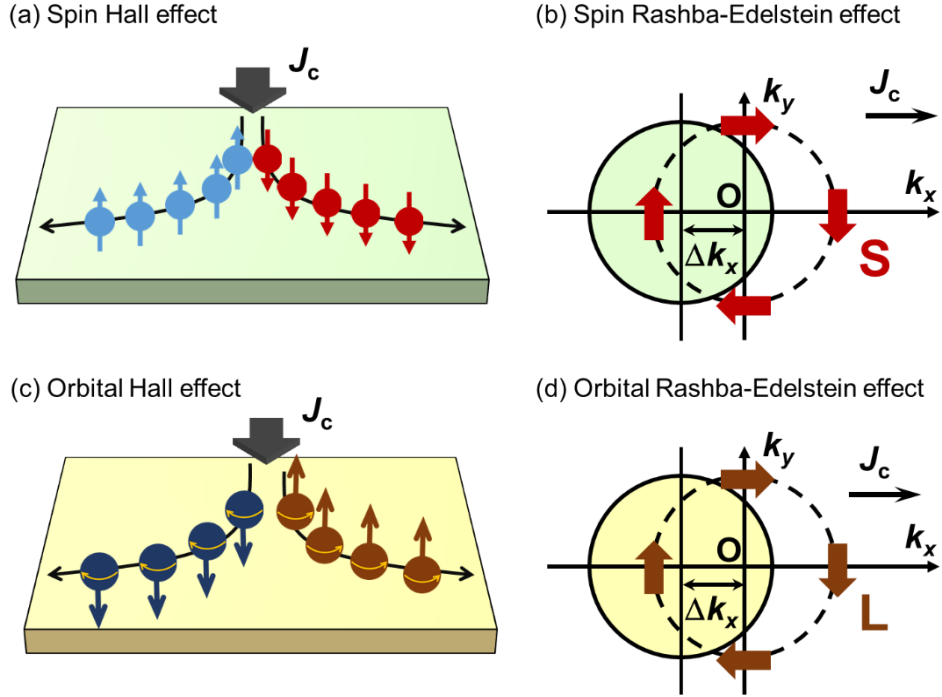

FIG. 1. Illustrations of the (a) spin Hall effect (SHE), (b) spin Rashba-Edelstein effect (SREE), (c) orbital Hall effect (OHE), and (d) orbital Rashba-Edelstein effect (OREE). The arrows in the SHE represent the spin polarizations, while those of the OHE are the orbital polarizations. The red arrows in REE represent the spin angular momentum (*S*) and the brown arrows in OREE represent the orbital angular momentum (*L*).

### III. Nitrogen treatment on Cu layer

A series of Ta(0.5)/Pt(3)/Co(1)/CuN$_x$(7) (units in nm) magnetic heterostructures, as depicted in Fig. 2(a), are prepared by magnetron sputtering. The Ta(0.5) layer serves as a seed layer to improve adhesion between the SiO$_2$ substrate and the Pt layer. The Pt/Co bilayer provides a robust interfacial PMA and the Cu capping layer is found to be beneficial for sustaining the PMA in Pt/Co system because of the immiscibility of Co and Cu [44,45]. The Cu thickness of 7 nm is chosen by realizing a suitable OOP coercivity [46]. All samples are deposited onto thermally oxidized Si substrates with a base pressure ~ $3 \times 10^{-8}$ Torr and the working pressure of Ar during deposition is set to be 3 mTorr. The $CuN_x$ layer is deposited by reactive sputtering through tuning the flow rates of Ar and $N_2$. The doping concentration $Q$ is defined as $Q = N_2$ flow rate / $(N_2 + Ar)$ flow rate, where the Ar flow rate is fixed at 30 sccm (standard cubic centimeter per minute) and the $N_2$ flow rates are set to be 2, 4, 6, 8, 10, and 12 sccm, which corresponds to $Q$ = 6.3, 11.8, 16.7, 21.1, 25, and 28.6%.

The structural properties of the deposited films are first characterized by the X-ray diffraction (XRD). The XRD spectrums of the films with $Q$ = 0%, 11.8%, and 25% are shown in Fig. 2(b). The strong peaks of Pt (111) and Cu (111) indicate that the films are well-textured. As $Q$ increases, the Cu (111) peak shifts toward a lower diffraction angle, which verifies that the nitrogen is incorporated into Cu. Next, the films are patterned into Hall bar devices with lateral dimensions of $5 \times 60$ μm² as shown in Fig. 2(c). Electrical and magnetic properties of the heterostructures are further examined by four-point probe measurement and anomalous Hall effect (AHE) measurement. Fig. 2(d) summarizes the measured longitudinal resistances of the devices. The quasilinear increase of the device resistance vs. $Q$ verifies the nitridation of Cu [42]. A representative OOP hysteresis loop obtained from a Pt(3)/Co(1)/$CuN_x$(7) ($Q$ = 25%) device by AHE is shown in Fig. 2(e). The squareness of the hysteresis loop indicates a robust PMA with an OOP coercive field $H_c$ ~ 20 Oe. The $Q$-dependent $H_c$ and saturation magnetization ($M_s$) are summarized in Fig. 2(f). The reduction of $H_c$ and $M_s$ suggest that excessive nitridation could be detrimental to the Co layer. This degradation of $H_c$ and $M_s$ might be owing to the intermixing of Co and $CuN_x$, since the nitrogen atom could be incorporated onto the Co surface during deposition. Notably, the magnetic anisotropy of the Co layer becomes in-plane as $Q > 30$ %, therefore, we focus on the devices with $Q < 30\%$ for the following SOT characterizations and current-induced switching measurements.

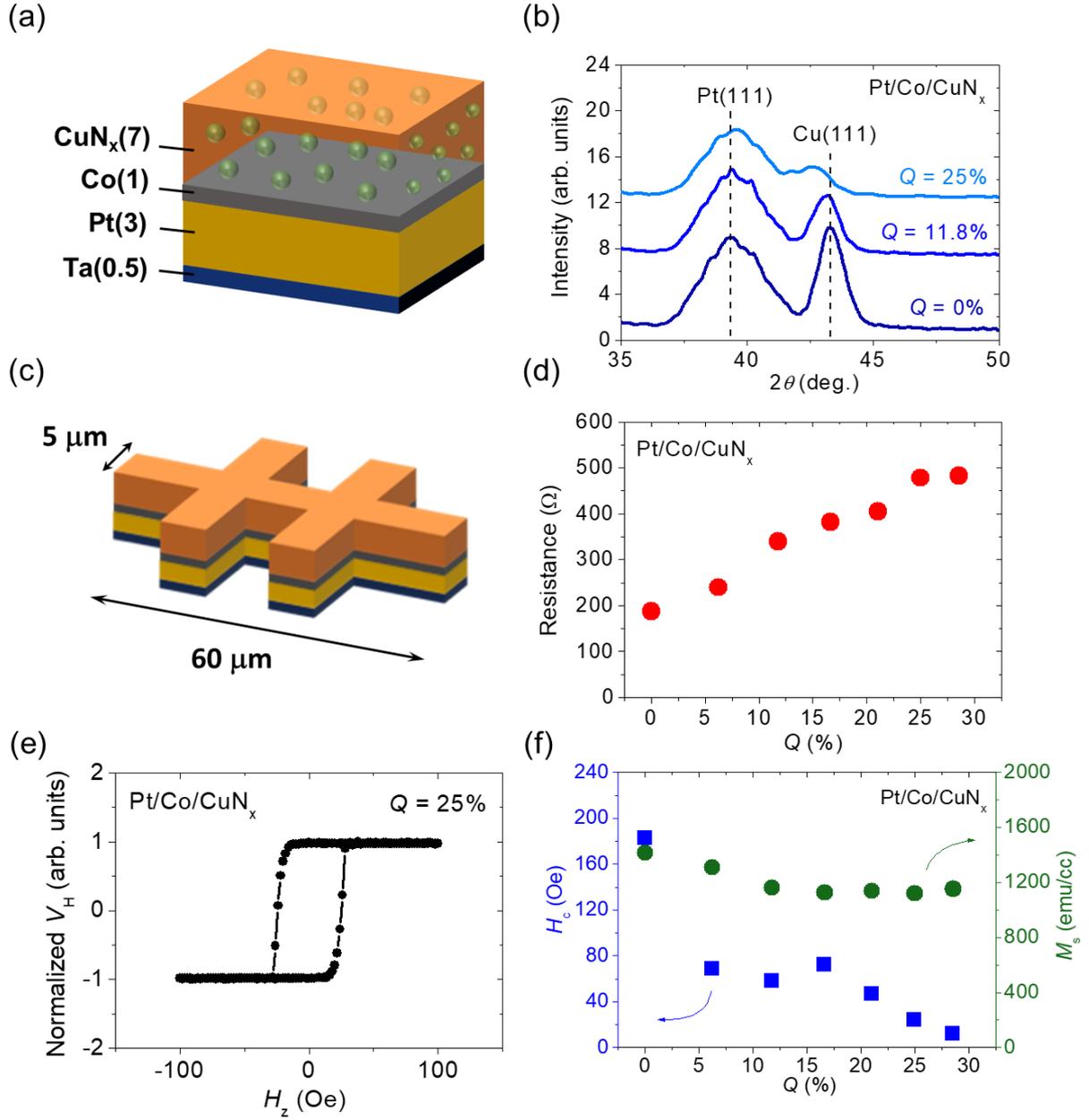

FIG. 2. (a) Illustration of the layer stack. (b) XRD patterns of the films with $Q = 0\%$, 11.8%, and 25%. (c) Illustration of a Hall bar device. (d) Longitudinal resistance of Pt(3)/Co(1)/CuN$_x$(7) devices as a function of $Q$. (e) Representative OOP hysteresis loop from a Pt(3)/Co(1)/CuN$_x$(7) ($Q = 25\%$) device. (f) OOP coercivity and saturation magnetization vs. $Q$ of Pt(3)/Co(1)/CuN$_x$(7) devices.

## IV. SOT characterization

To quantify the SOT efficiency of Pt/Co/CuN$_x$ devices, we perform current-induced hysteresis loop shift measurement [47] as illustrated in Fig. 3(a). A dc current ($I_{dc}$) supplied by Keithley 2400 source meter is injected into the current channel (longitudinal, $x$ direction), and the Hall voltage ($V_H$) is detected by Keithley 2000 multimeter along the transverse $y$ direction. With the biasing in-plane magnetic fields ($H_x$), the domain wall moments are realigned and the Dzyaloshinskii-Moriya interaction effective field ($H_{DMI}$) is overcome, which leads to damping-like SOT-driven domain expansion and domain wall propagation [48,49]. The damping-like SOT acting on the magnetization then manifests as an OOP effective field $H_z^{eff}$. Fig. 3(b) shows representative hysteresis loop shifts from a Pt/Co/CuN$_x$ ($Q$ = 25%) device under $I_{dc}$ = ±2.5 mA and $H_x$ = 2.5 kOe. Current-induced effective fields are summarized in Fig. 3(c). It is noted that a positive slope ($H_z^{eff}/I_{dc}$) is obtained under a positive external in-plane field, which clarifies the overall SOT acting on ferromagnet Co has a positive sign. The SOT efficacy in terms of effective field per current density, $\chi = H_z^{eff}/J_{dc}$, as a function of $H_x$ is shown in Fig. 3(d) [46], where it saturates at $\chi$ ~ 70 Oe m$^2$/10$^{11}$ A with $H_{DMI}$ ~ 1.5 kOe. It is worth noting that the SOT efficacy of Pt/Co/CuN$_x$ is greater than that of the Pt/Co/MgO standard (control) sample but comes with a smaller DMI field [46]. This implies that the CuN$_x$ layer generates a negative SOT and the enhanced current-induced SOT might not come from the strong SOC at the interface, which further suggests that the CuN$_x$-contributed SOT efficacy is of a SOC-free origin.

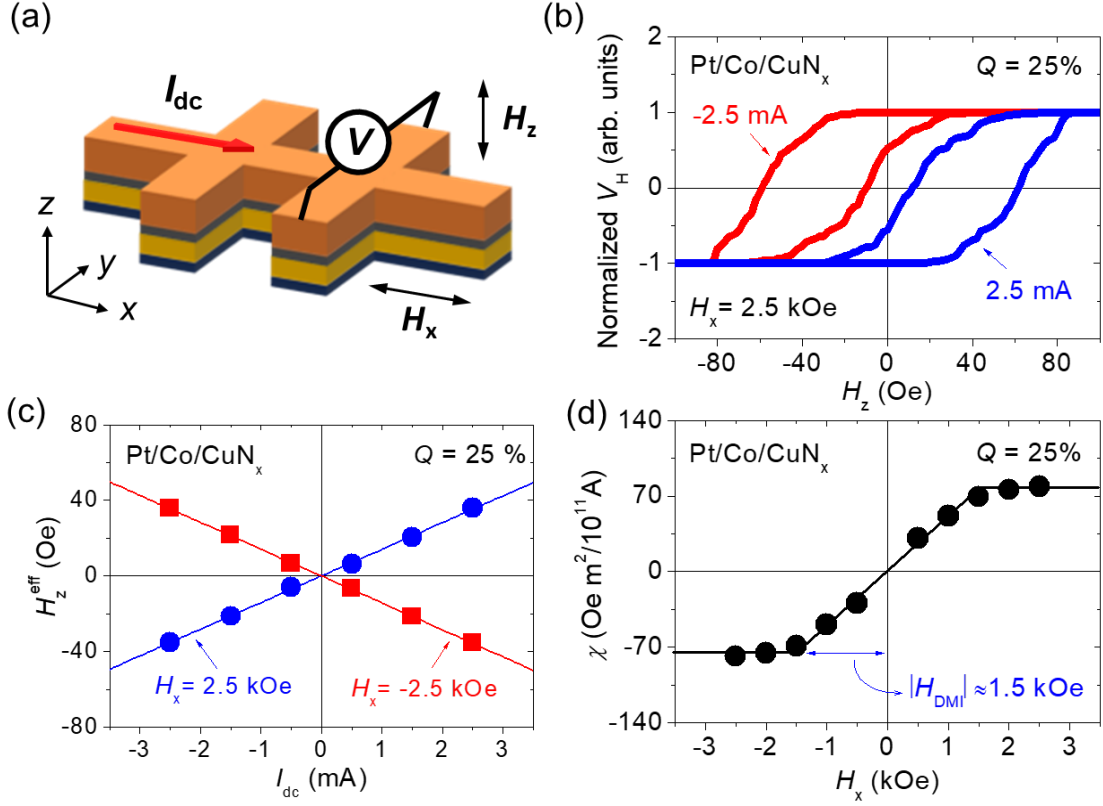

FIG. 3. Hysteresis loop shift measurement on a Pt(3)/Co(1)/CuN$_x$(7) ($Q$ = 25%) device. (a) Illustration of the measurement configuration. (b) Representative hysteresis loop shifts under $I$ = ±2.5 mA and $H_x$ = 2.5 kOe. (c) Current-induced effective fields as a function of applied dc current under $H_x$ = ±2.5 kOe. (d) The SOT efficacy $\chi$ as a function of in-plane field $H_x$.

The damping-like SOT efficiency is then evaluated from $\chi$ through [47]

$$\xi_{DL} = (2e/\hbar)(2/\pi)\mu_0 M_s t_{Co} w t_{SOT} \chi,$$

where $t_{SOT} = t_{Pt} + t_{CuN}$ = 10 nm is the overall SOT layer thickness. Figure 4(a) and (b) show the SOT efficacy $\chi$ and the damping-like SOT efficiency $\xi_{DL}$ as functions of Q. The dashed line in Fig. 4(b) represents the efficiency $\xi_{DL}$ ~ 0.12 of the Pt(3)/Co(1)/MgO(2)/Ta(2) control sample [46]. When $Q$ < 11.8 %, the multilayer suffers from the shunting effect (into the Cu layer) and results in a $\xi_{DL}$ smaller than the control case. $\xi_{DL}$ gets enhanced with increasing $Q$ and surpasses the pure

Pt value when $Q > 11.8\%$. Based on the SOT enhancement and the layer structure geometry, the SOT generation from the Co/CuN$_x$ portion has a sign opposite to that from the bottom Pt. The effective SOT conductivity ($\sigma_{SOT}$) of Pt/Co/CuN$_x$ reaches $\sigma_{SOT} = \xi_{DL}/\rho_{SOT} = 4.7 \times 10^5$ $\Omega^{-1}$m$^{-1}$, which is two times larger than that of the Pt/Co/MgO control sample and much greater than the W-based [50] and Ta-based magnetic heterostructures [42,46]. Note that the SOT-contributing layer is the Pt and the CuN$_x$ layers combined, and $\rho_{SOT}$ considers the resistivities from both Pt and CuN$_x$ [46].

The enhanced SOT conductivity observed in the Pt/Co/CuN$_x$ devices can be attributed to two major SOT contributions, namely the SHE-induced spin current from the Pt bottom layer and the orbital current from the CuN$_x$ capping layer. It is well known that Pt is a HM with a positive and sizable spin Hall ratio that can effectively generate a transverse spin current, as illustrated in Fig. 4(c) (spin current is the dominating mechanism in Pt). As for the CuN$_x$ layer, pristine Cu is known for its minimal SOC as mentioned above. However, a finite OHE and/or OREE can still be generated in Cu when the energy of the *d* state is much lower than the Fermi level [23]. Furthermore, the introduction of oxygen or nitrogen induces *p-d* hybridization and the energy of the *d* state can be promoted to near the Fermi level [17]. Consequently, the OAM-induced OHE or OREE is enhanced. Thus, the SHE contributed SOT is negligible and the OAM-induced transverse orbital current is responsible for the SOT production from the CuN$_x$ layer, as depicted in Fig. 4(d). Notably, one previous report revealed that a CuO layer can generate SOT with positive sign on the adjacent Py (Ni$_{80}$Fe$_{20}$) layer [13]; however, the experimental results from our loop shift measurement indicate that the CuN$_x$ exerts a *negative-sign* torque on the Co layer. We speculate the torque from CuN$_x$ is induced from first the OAM and then affected by the negative *L-S* conversion coefficient $\eta$ in the Co/CuN$_x$ system [46]. This feature serves as an unambiguous evidence of the OAM-induced torque in Co/CuN$_x$ since the intrinsic SHE of Cu should be positive and nitridation will only affect the magnitude of the SHE-SOT but not the sign. Theoretical study with *ab initio* calculations also supported this conclusion [51]. The existence of the orbital current from $CuN_x$ is further confirmed by additional measurements on samples with various Pt thicknesses and layer structure configurations [46]. Although it has been reported that the OREE dominates the orbital torque generation in CuO, in which the CuO/ferromagnet interface plays an essential role [14,18], following studies suggest that the bulk OHE could also contribute significantly to the orbital torque because the oxygen or nitrogen can be incorporated into the Cu layer and thus resulting in the localized OAM effect [19]. Accordingly, we suspect that both the interfacial OREE (at the $CuN_x$/Co interface) and the bulk OHE (from the $CuN_x$ layer) can contribute to the generation of the observed orbital current.

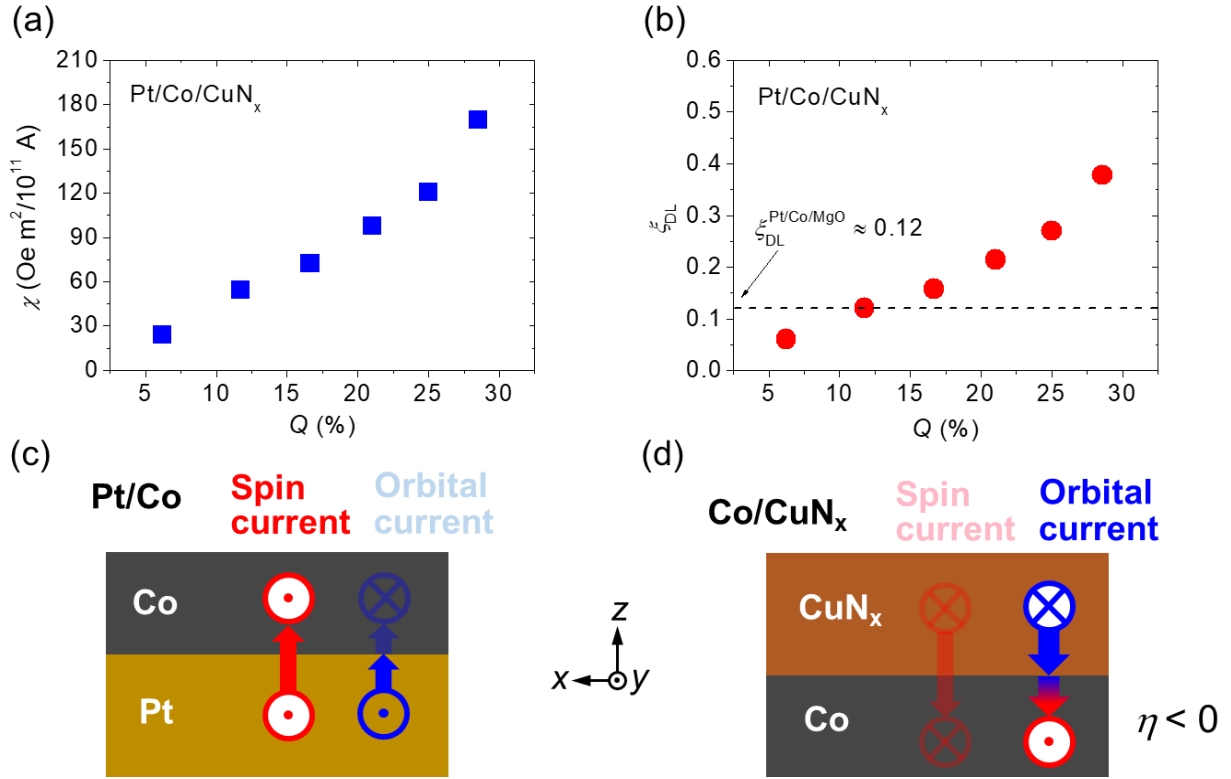

FIG. 4. (a) SOT efficacy $\chi$ and (b) SOT efficiency $\xi_{DL}$ as functions of $Q$. Illustrations of SOT generations from the SHE and the OHE in (c) the Pt/Co bilayer (spin current dominates) (d) the Co/$CuN_x$ bilayer (orbital current dominates). The orbital polarization in the $CuN_x$ is pointed along -$y$ direction because the structure geometry is opposite to the Pt/Co.

**Neuromorphic SOT switching**

After confirming the sizable SOT efficiency in Pt/Co/CuN$_x$ devices, we perform current-induced SOT switching measurements on the same devices. SOT-induced switching has been shown to have memristive behavior and the feasibility of neuromorphic computing [27-31]. To verify the same applicability in devices with orbital currents, we examine the switching behavior variation with respect to the nitrogen concentration in the CuN$_x$ layer. The measurement utilizes a pulse-current with pulse width $t_{\text{pulse}} = 50$ ms injecting into the current channel and applying an external in-plane magnetic field to overcome $H_{\text{DMI}}$ [48], as shown in Fig. 5a. The Co magnetization state is initialized to magnetization-down by a negative current larger than the critical switching current before executing memristive switching. The measurement sequence is followed by sweeping the pulse current $-I_{\max} \to 0 \to I_{\max} \to 0 \to -I_{\max}$ until the $I_{\max}$ is large enough to reach full switching. Fig. 5(b) shows the representative $V_H$-$I_{\max}$ loops with $I_{\max}$ ranges from 3 mA to 11 mA, as obtained from a Pt/Co/CuN$_x$ ($Q = 25\%$) device. The intermediate states observed here reveal that the memristive behavior can be operated by the orbital-enhanced SOT. The variation of such memristive switching behavior from devices with $Q = 6.3\%$, 11.8%, and 25% is shown in Fig. 5(c). The memristive window ($\Delta J_{\max}^{\text{window}}$) of $Q = 25\%$ is marked as the light blue area in Fig. 5(c), which is defined as the range between the maximum (90% of the full switching) and the minimum (1% of the full switching) current densities to obtain intermediate states. The switching current density at which 90% of the full switching takes place is denoted as the critical switching current density ($J_c$). $J_c$ of each device is summarized in Fig. 5(d), where it decreases as increasing the doping concentration $Q$. This again verifies that the SOT efficiency can be effectively enhanced by the orbital current from CuN$_x$. Notably, the critical switching current density achieves as low as $J_c \sim 5 \times 10^{10}$ A/m$^2$ (for $Q = 28.6\%$). Compared to the Pt control sample, the switching efficiency is considerably improved by the coexistence of the spin current and the orbital current [46].

The memristive windows for devices with different $Q$s are then summarized in Fig. 5(e). It shows that the memristive window shrinks with increasing $Q$, which hints that the memristive window is a tradeoff for the switching efficiency, as further confirm in Fig. 5(f). If one wants to obtain a wider memristive window, a larger $J_c$ is required to fully control the magnetization. This observation poses a potential challenge for the spin-orbitronic neuromorphic devices development. Besides the issue we address here, recent investigations on SOT-driven neuromorphic switching have also raised some unsolved issues: First, the size of the memristive window will also reduce significantly as the ambient temperature rises [28]. Second, the memristive switching that relies on domain wall propagation will be constrained when the device downscales to nano-size and becomes single domain [52]. A recent study demonstrated that neuromorphic switching can be tentatively operated by domain nucleation rather than by domain wall motion [27], which should be a possible solution for this size issue.

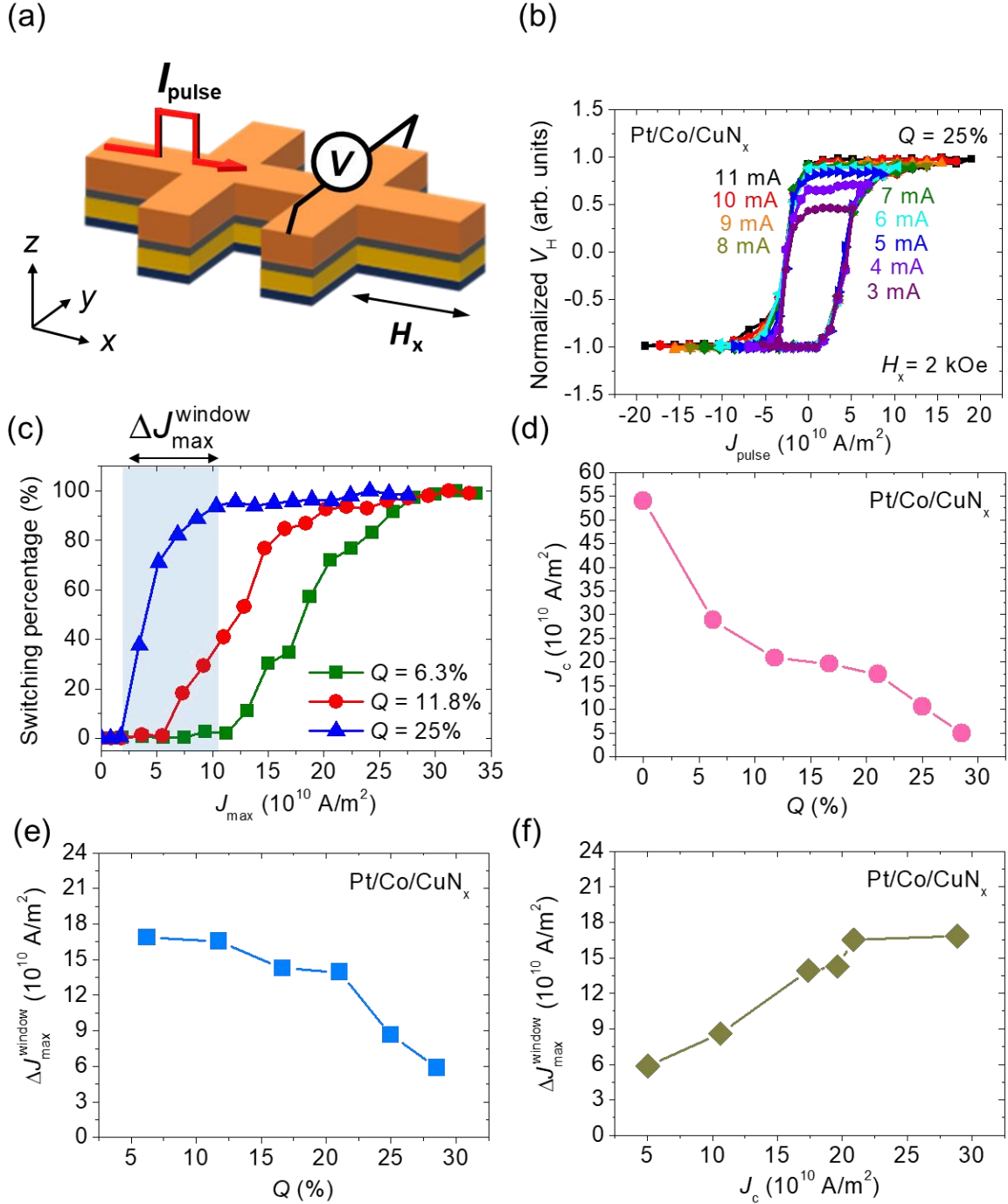

FIG. 5. Current-induced SOT-driven memristive switching. (a) Illustration of the switching measurement. (b) Switching loops with various maximum applied current densities $J_{\text{pulse}}$. (c) Switching percentage as functions of maximum applied current densities $J_{\text{max}}$ for devices with $Q$ = 6.3%, 11.8%, and 25%. (d) Critical switching current density ($J_c$) as a function of $Q$. (e)

Memristive switching window as a function of $Q$. (f) Memristive switching window as a function of $J_c$.

**Conclusion**

In conclusion, we systematically analyze the influence of nitridation on magnetic properties, SOT efficiencies, and neuromorphic switching behavior in Pt/Co/CuN$_x$ magnetic heterostructures. The CuN$_x$ capping layer is found to generate a sizable orbital current. The effective damping-like SOT efficiency can be tailored by controlling the nitrogen doping concentration, and reaches $\xi_{DL}$ ~ 0.4 when $Q$ = 28.6%. Notably, the effective SOT conductivity achieves $\sigma_{SOT} = 4.7 \times 10^5$ $\Omega^{-1}m^{-1}$ in Pt/Co/CuN$_x$, which is two times larger than the Pt/Co/MgO control sample. The SOT switching measurement further verifies the perpendicular magnetization can be manipulated by a low critical switching current density $J_c$ ~ $5 \times 10^{10}$ A/m$^2$. Last but not least, the neuromorphic switching behavior is successfully tuned by the orbital current. However, the tunability of the SOT-driven neuromorphic switching, as quantified by the memristive switching window, is found to be diminished as increasing the nitrogen concentration in Cu. It is therefore critical to note that enhancing the SOT efficiency for a better standard SOT MRAM performance is at the sacrifice of its mouldability for neuromorphic computing applications.

**Author contribution**

Tian-Yue Chen conceived the experiments and drafted the manuscript. Yu-Chan Hsiao prepared the samples and performed the hysteresis loop shift measurement and current-induced SOT switching measurement. Wei-Bang Liao performed the analysis of the data and the XRD measurement. Chi-Feng Pai proposed and supervised the study.

**Competing interests**

Authors declare no competing interests.

**Acknowledgement**

This work is supported by the Ministry of Science and Technology of Taiwan (MOST) under grant No. MOST-110-2636-M-002 -013 and by the Center of Atomic Initiative for New Materials (AI-Mat) and the Advanced Research Center of Green Materials Science and Technology, National Taiwan University from the Featured Areas Research Center Program within the framework of the Higher Education Sprout Project by the Ministry of Education (MOE) in Taiwan under grant No. NTU-110L9008.

[*] Email: cfpai@ntu.edu.tw

**References**

[1] J. E. Hirsch, Spin Hall Effect, Phys. Rev. Lett. **83**, 1834 (1999).

[2] R. Karplus and J. M. Luttinger, Hall Effect in Ferromagnetics, Phys. Rev. **95**, 1154 (1954).

[3] M. I. Dyakonov and V. I. Perel, Current-Induced Spin Orientation of Electrons in Semiconductors, Phys. Lett. A **35**, 459 (1971).

[4] A. Hoffmann, Spin Hall Effects in Metals, IEEE Trans. Magn. **49**, 5172 (2013).

[5] L. Liu, C.-F. Pai, Y. Li, H. W. Tseng, D. C. Ralph, and R. A. Buhrman, Spin-Torque Switching with the Giant Spin Hall Effect of Tantalum, Science **336**, 555 (2012).

[6] C.-F. Pai, L. Liu, Y. Li, H. W. Tseng, D. C. Ralph, and R. A. Buhrman, Spin Transfer Torque Devices Utilizing the Giant Spin Hall Effect of Tungsten, Appl. Phys. Lett. **101**, 122404 (2012).

[7] C.-F. Pai, Y. Ou, L. H. Vilela-Leão, D. C. Ralph, and R. A. Buhrman, Dependence of the Efficiency of Spin Hall Torque on the Transparency of Pt/Ferromagnetic Layer Interfaces, Phys. Rev. B **92**, 064426 (2015).

[8] V. M. Edelstein, Spin Polarization of Conduction Electrons Induced by Electric Current in Two-Dimensional Asymmetric Electron Systems, Solid State Commun. **73**, 233 (1990).

[9] J. C. R. Sánchez, L. Vila, G. Desfonds, S. Gambarelli, J. P. Attané, J. M. De Teresa, C. Magén, and A. Fert, Spin-to-Charge Conversion Using Rashba Coupling at the Interface between

Non-Magnetic Materials, Nat. Commun. **4**, 2944 (2013).

[10] I. M. Miron, G. Gaudin, S. Auffret, B. Rodmacq, A. Schuhl, S. Pizzini, J. Vogel, and P. Gambardella, Current-Driven Spin Torque Induced by the Rashba Effect in a Ferromagnetic Metal Layer, Nat. Mater. **9**, 230 (2010).

[11] Z. Chi, Y.-C. Lau, X. Xu, T. Ohkubo, K. Hono, and M. Hayashi, The Spin Hall Effect of Bi-Sb Alloys Driven by Thermally Excited Dirac-Like Electrons, Sci. Adv. **6**, eaay2324 (2020).

[12] T. Y. Chen, C. W. Peng, T. Y. Tsai, W. B. Liao, C. T. Wu, H. W. Yen, and C. F. Pai, Efficient Spin-Orbit Torque Switching with Nonepitaxial Chalcogenide Heterostructures, ACS Appl. Mater. Interfaces **12**, 7788 (2020).

[13] H. An, Y. Kageyama, Y. Kanno, N. Enishi, and K. Ando, Spin-Torque Generator Engineered by Natural Oxidation of Cu, Nat. Commun. **7**, 13069 (2016).

[14] Y. Kageyama, Y. Tazaki, H. An, T. Harumoto, T. Gao, J. Shi, and K. Ando, Spin-Orbit Torque Manipulated by Fine-Tuning of Oxygen-Induced Orbital Hybridization, Sci. Adv. **5**, eaax4278 (2019).

[15] T. Tanaka, H. Kontani, M. Naito, T. Naito, D. S. Hirashima, K. Yamada, and J. Inoue, Intrinsic Spin Hall Effect and Orbital Hall Effect in 4d and 5d Transition Metals, Phys. Rev. B **77**, 165117 (2008).

[16] D. Lee, D. Go, H. J. Park, W. Jeong, H. W. Ko, D. Yun, D. Jo, S. Lee, G. Go, J. H. Oh *et al.*, Orbital Torque in Magnetic Bilayers, Nat. Commun. **12**, 6710 (2021).

[17] D. Go, D. Jo, T. Gao, K. Ando, S. Blügel, H.-W. Lee, and Y. Mokrousov, Orbital Rashba Effect in a Surface-Oxidized Cu Film, Phys. Rev. B **103**, L121113 (2021).

[18] S. Ding, A. Ross, D. Go, L. Baldrati, Z. Ren, F. Freimuth, S. Becker, F. Kammerbauer, J. Yang, G. Jakob *et al.*, Harnessing Orbital-to-Spin Conversion of Interfacial Orbital Currents for Efficient Spin-Orbit Torques, Phys. Rev. Lett. **125**, 177201 (2020).

[19] J. Kim, D. Go, H. Tsai, D. Jo, K. Kondou, H.-W. Lee, and Y. Otani, Nontrivial Torque Generation by Orbital Angular Momentum Injection in Ferromagnetic-Metal/Cu/Al$_2$O$_3$ Trilayers,

Phys. Rev. B **103**, L020407 (2021).

[20] S. Lee, M.-G. Kang, D. Go, D. Kim, J.-H. Kang, T. Lee, G.-H. Lee, J. Kang, N. J. Lee, Y. Mokrousov *et al.*, Efficient Conversion of Orbital Hall Current to Spin Current for Spin-Orbit Torque Switching, Commun. Phys. **4**, 234 (2021).

[21] D. Go, D. Jo, C. Kim, and H.-W. Lee, Intrinsic Spin and Orbital Hall Effects from Orbital Texture, Phys. Rev. Lett. **121**, 086602 (2018).

[22] Y.-G. Choi, D. Jo, K.-H. Ko, D. Go, K.-H. Kim, H. Gyum Park, C. Kim, B.-C. Min, G.-M. Choi, and H.-W. Lee, Observation of the Orbital Hall Effect in a Light Metal Ti, arXiv e-prints, arXiv:2109.14847 (2021).

[23] D. Jo, D. Go, and H.-W. Lee, Gigantic Intrinsic Orbital Hall Effects in Weakly Spin-Orbit Coupled Metals, Phys. Rev. B **98**, 214405 (2018).

[24] D. Go and H.-W. Lee, Orbital Torque: Torque Generation by Orbital Current Injection, Phys. Rev. Res. **2**, 013177 (2020).

[25] T. C. Chuang, C. F. Pai, and S. Y. Huang, Cr-Induced Perpendicular Magnetic Anisotropy and Field-Free Spin-Orbit-Torque Switching, Phys. Rev. Appl. **11**, 061005 (2019).

[26] T.-Y. Chen, H.-I. Chan, W.-B. Liao, and C.-F. Pai, Current-Induced Spin-Orbit Torque and Field-Free Switching in Mo-Based Magnetic Heterostructures, Phys. Rev. Appl. **10**, 044038 (2018).

[27] J. Zhou, T. Zhao, X. Shu, L. Liu, W. Lin, S. Chen, S. Shi, X. Yan, X. Liu, and J. Chen, Spin–Orbit Torque-Induced Domain Nucleation for Neuromorphic Computing, Adv. Mater. **33**, 2103672 (2021).

[28] W.-B. Liao, T.-Y. Chen, Y.-C. Hsiao, and C.-F. Pai, Pulse-Width and Temperature Dependence of Memristive Spin–Orbit Torque Switching, Appl. Phys. Lett. **117**, 182402 (2020).

[29] S. Fukami, C. Zhang, S. DuttaGupta, A. Kurenkov, and H. Ohno, Magnetization Switching by Spin-Orbit Torque in an Antiferromagnet-Ferromagnet Bilayer System, Nat. Mater. **15**, 535 (2016).

[30] G. J. Lim, W. L. Gan, W. C. Law, C. Murapaka, and W. S. Lew, Spin-Orbit Torque Induced Multi-State Magnetization Switching in Co/Pt Hall Cross Structures at Elevated Temperatures, J. Mag. Mag. Mater. **514**, 167201 (2020).

[31] S. Zhang, S. Luo, N. Xu, Q. Zou, M. Song, J. Yun, Q. Luo, Z. Guo, R. Li, W. Tian *et al.*, A Spin–Orbit-Torque Memristive Device, Adv. Electron. Mater. **5**, 1800782 (2019).

[32] H. Mulaosmanovic, E. Chicca, M. Bertele, T. Mikolajick, and S. Slesazeck, Mimicking Biological Neurons with a Nanoscale Ferroelectric Transistor, Nanoscale **10**, 21755 (2018).

[33] S. Boyn, J. Grollier, G. Lecerf, B. Xu, N. Locatelli, S. Fusil, S. Girod, C. Carrétéro, K. Garcia, S. Xavier *et al.*, Learning through Ferroelectric Domain Dynamics in Solid-State Synapses, Nat. Commun. **8**, 14736 (2017).

[34] D. Kuzum, R. G. D. Jeyasingh, B. Lee, and H. S. P. Wong, Nanoelectronic Programmable Synapses Based on Phase Change Materials for Brain-Inspired Computing, Nano Letters **12**, 2179 (2012).

[35] S. Ambrogio, N. Ciocchini, M. Laudato, V. Milo, A. Pirovano, P. Fantini, and D. Ielmini, Unsupervised Learning by Spike Timing Dependent Plasticity in Phase Change Memory (Pcm) Synapses, Front. Neurosci. **10**, 56 (2016).

[36] C. Sung, H. Hwang, and I. K. Yoo, Perspective: A Review on Memristive Hardware for Neuromorphic Computation, J. Appl. Phys. **124**, 151903 (2018).

[37] Y. Zhang, P. Huang, B. Gao, J. Kang, and H. Wu, Oxide-Based Filamentary Rram for Deep Learning, J. Phys. D: Appl. Phys **54**, 083002 (2020).

[38] K. Y. Camsari, B. M. Sutton, and S. Datta, P-Bits for Probabilistic Spin Logic, Appl. Phys. Rev. **6**, 011305 (2019).

[39] M. Romera, P. Talatchian, S. Tsunegi, F. Abreu Araujo, V. Cros, P. Bortolotti, J. Trastoy, K. Yakushiji, A. Fukushima, H. Kubota *et al.*, Vowel Recognition with Four Coupled Spin-Torque Nano-Oscillators, Nature **563**, 230 (2018).

[40] I. Chakraborty, A. Jaiswal, A. K. Saha, S. K. Gupta, and K. Roy, Pathways to Efficient

Neuromorphic Computing with Non-Volatile Memory Technologies, Appl. Phys. Rev. **7**, 021308 (2020).

[41] Y. Niimi, Y. Kawanishi, D. H. Wei, C. Deranlot, H. X. Yang, M. Chshiev, T. Valet, A. Fert, and Y. Otani, Giant Spin Hall Effect Induced by Skew Scattering from Bismuth Impurities inside Thin Film CuBi Alloys, Phys. Rev. Lett. **109**, 156602 (2012).

[42] T.-Y. Chen, C.-T. Wu, H.-W. Yen, and C.-F. Pai, Tunable Spin-Orbit Torque in Cu-Ta Binary Alloy Heterostructures, Phys. Rev. B **96**, 104434 (2017).

[43] Z. Xu, G. D. H. Wong, J. Tang, E. Liu, W. Gan, F. Xu, and W. S. Lew, Large Spin Hall Angle Enhanced by Nitrogen Incorporation in Pt Films, Appl. Phys. Lett. **118**, 062406 (2021).

[44] H. K. Gweon, S. J. Yun, and S. H. Lim, A Very Large Perpendicular Magnetic Anisotropy in Pt/Co/MgO Trilayers Fabricated by Controlling the MgO Sputtering Power and Its Thickness, Sci. Rep. **8**, 1266 (2018).

[45] S. Bandiera, R. C. Sousa, B. Rodmacq, and B. Dieny, Asymmetric Interfacial Perpendicular Magnetic Anisotropy in Pt/Co/Pt Trilayers, IEEE Magn. Lett. **2**, 3000504 (2011).

[46] See Supplemental Material at [URL] for details of Pt/Co/Cu Optimization, Resistivity, Control Sample, (Spin Plus Orbital) SOT Conductivity, Intrinsic Spin Orbit Torque of Co, Orbital Current from $CuN_x$, Neuromorphic Switching from Control Sample. .

[47] C.-F. Pai, M. Mann, A. J. Tan, and G. S. D. Beach, Determination of Spin Torque Efficiencies in Heterostructures with Perpendicular Magnetic Anisotropy, Phys. Rev. B **93**, 144409 (2016).

[48] O. J. Lee, L. Q. Liu, C. F. Pai, Y. Li, H. W. Tseng, P. G. Gowtham, J. P. Park, D. C. Ralph, and R. A. Buhrman, Central Role of Domain Wall Depinning for Perpendicular Magnetization Switching Driven by Spin Torque from the Spin Hall Effect, Phys. Rev. B **89**, 024418 (2014).

[49] S. Emori, E. Martinez, K.-J. Lee, H.-W. Lee, U. Bauer, S.-M. Ahn, P. Agrawal, D. C. Bono, and G. S. D. Beach, Spin Hall Torque Magnetometry of Dzyaloshinskii Domain Walls, Phys. Rev. B **90**, 184427 (2014).

[50] W.-B. Liao, T.-Y. Chen, Y. Ferrante, S. S. P. Parkin, and C.-F. Pai, Current-Induced Magnetization Switching by the High Spin Hall Conductivity α-W, Phys. Status Solidi RRL **13**, 1900408 (2019).

[51] M. Gradhand, D. V. Fedorov, P. Zahn, and I. Mertig, Spin Hall Angle Versus Spin Diffusion Length: Tailored by Impurities, Phys. Rev. B **81**, 245109 (2010).

[52] A. Kurenkov, C. Zhang, S. DuttaGupta, S. Fukami, and H. Ohno, Device-Size Dependence of Field-Free Spin-Orbit Torque Induced Magnetization Switching in Antiferromagnet/Ferromagnet Structures, Appl. Phys. Lett. **110**, 092410 (2017).

# Supplemental Material

# Tailoring neuromorphic switching by CuN$_x$ mediated orbital currents

Tian-Yue Chen[1], Yu-Chan Hsiao[1], Wei-Bang Liao[1], and Chi-Feng Pai[1,2*]

[1]*Department of Materials Science and Engineering, National Taiwan University, Taipei 10617, Taiwan*

[2]*Center of Atomic Initiative for New Materials, National Taiwan University, Taipei 10617, Taiwan*

**Supplementary Note 1. Pt/Co/Cu optimization**

**Supplementary Note 2. Layer resistivity characterizations**

**Supplementary Note 3. Pt/Co/MgO/Ta control sample**

**Supplementary Note 4. The (spin plus orbital) SOT conductivity**

**Supplementary Note 5. Intrinsic spin-orbit torque from the Co layer**

**Supplementary Note 6. Orbital currents from the CuN$_x$ layer**

**Supplementary Note 7. Neuromorphic switching from the control sample**

## Supplementary Note 1. Pt/Co/Cu optimization

It has been reported that the Cu capping layer is beneficial for enhancing perpendicular magnetic anisotropy (PMA) in the Pt/Co system [44,45]. To get a better PMA and an optimized $H_c$, we prepare a series of samples, namely Pt(3)/Co(1)/Cu($t$) and perform magneto-optical Kerr effect (MOKE) measurement. A representative out-of-plane (OOP) hysteresis loop from Pt(3)/Co(1)/Cu(7) is shown in Fig. S1(a). Figure S1(b) shows the summarized OOP coercivity as a function of Cu thickness. The coercivity increases with the increasing Cu then saturates at $t \sim 4.5$ nm with $H_c \sim 200$ Oe. Considering the possibility of resistance engineering from nitridation, the optimized Pt(3)/Co(1)/Cu(7) is chosen to conduct the following experiments.

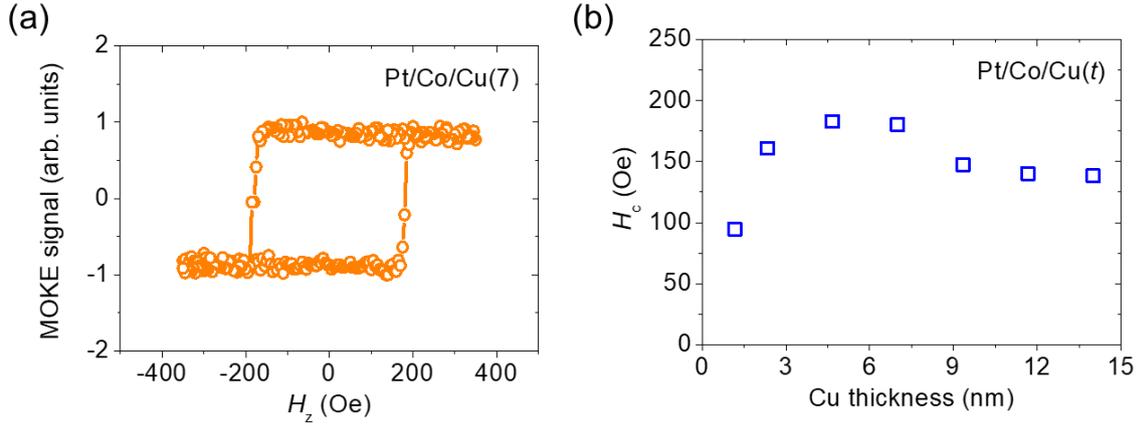

FIG. S1. (a) OOP hysteresis loop from a Pt(3)/Co(1)/Cu(7) sample. (b) OOP coercivity as a function of Cu thickness.

## Supplementary Note 2. Layer resistivity characterizations

To evaluate the current distribution in Pt(3)/Co(1)/CuN$_x$(7) trilayer, we construct a simple bilayer model as illustrated in Fig. S2(a). The effective SOT layer is 10 nm thick, which consists of the 3 nm Pt and the 7 nm CuN$_x$ layers. The resistivity of Co is estimated to be $\rho_{Co} \sim 50$ u$\Omega$-cm by the inverted resistance versus thickness as shown in Fig. S2(b). The measured resistance can be expressed as $R^{-1} = R_{SOT}^{-1} + R_{Co}^{-1}$, and the $R_{Co}^{-1} = A/(L\rho_{Co})$, where $L = 30$ µm, $A = wt = 5 \times 10^{-15}$ m$^2$. The $\rho_{SOT}$ of each $Q$ is summarized in Fig.S2(c). The resistivity of the CuN$_x$ layer alone is further summarized in Fig. S2(d).

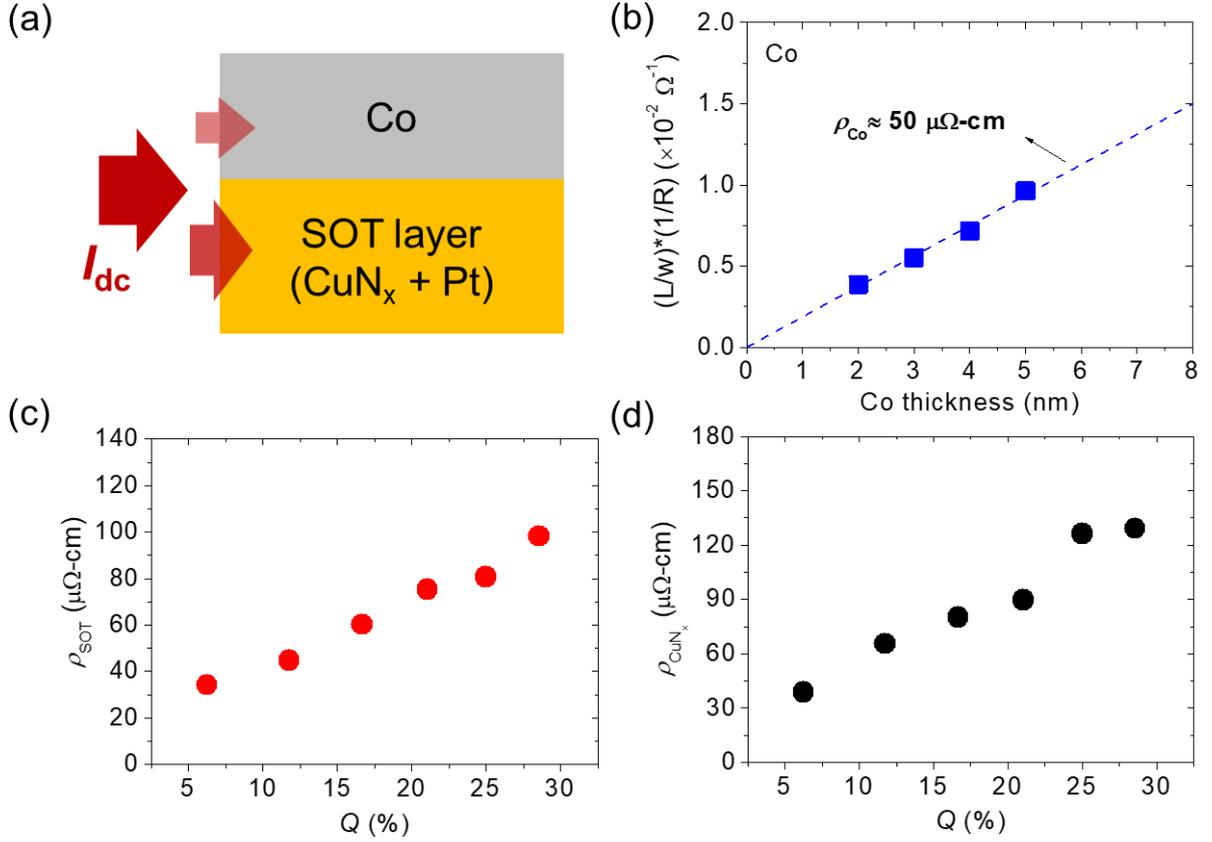

FIG. S2. (a) Illustration of bilayer model. (b) Inverse resistance of Co. (c) Resistivity of the effective SOT layer as a function of $Q$. (d) Resistivity of the $CuN_x$ layer with respect to $Q$.

## Supplementary Note 3. Pt/Co/MgO/Ta control sample

The OOP hysteresis loop of the control sample Pt(3)/Co(1)/MgO(2)/Ta(2) is shown in Fig. S3(a) with $H_c \sim 200$ Oe (measured through the AHE using a Hall bar device). Representative hysteresis loop shifts are shown in Fig. S3(b) under $I = \pm 2.5$ mA. The current-induced effective field as a function of applied current is summarized in Fig. S3(c). The effective field per current density (the efficacy) saturates at $\chi \sim 44$ Oe m$^2$/10$^{11}$ A with a DMI effective field $\sim 4000$ Oe, as shown in Fig. S3(d). The damping-like SOT efficiency is evaluated to be $\xi_{DL} \sim 0.12$.

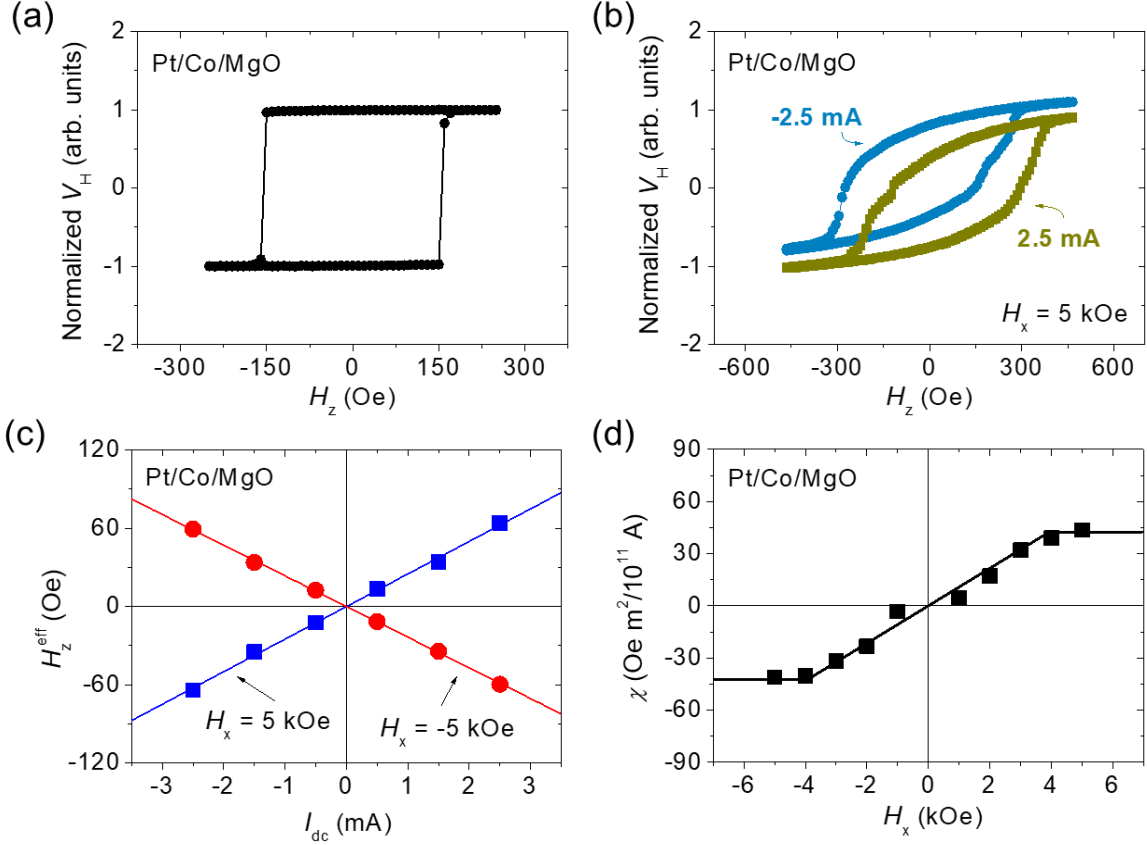

FIG. S3. Hysteresis loop shift measurement on a Pt(3)/Co(1)/MgO(2)/Ta(2) standard device. (a) OOP AHE loop. (b) Representative current-induced shifted hysteresis loops. (c) Current-induced effective fields. (d) The SOT efficacy as a function of external field along *x*-direction.

## Supplementary Note 4. The (spin plus orbital) SOT conductivity

Besides the damping-like SOT efficiency, we further evaluate the "SOT conductivity" $\sigma_{SOT}$, which includes the spin Hall conductivity (SHC) and the orbital Hall conductivity (OHC), in Pt/Co/CuN$_x$. The SOT conductivity is defined as $\sigma_{SOT} = \xi_{DL}/\rho_{SOT}$ and it is a critical figure of merit to benchmark the current-to-torque conversion. As shown in Fig. S4(a), the resistivity-dependent damping-like SOT efficiency gives rise to a constant SOT conductivity of Pt/Co/CuN$_x$, $\sigma_{SOT}$ ~ 4.7 × 10$^5$ Ω$^{-1}$m$^{-1}$, which is much larger than the control sample case, $\sigma_{SOT}$ ~ 2.0 × 10$^5$ Ω$^{-1}$m$^{-1}$

In general, the linear resistivity dependence of the damping-like SOT efficiency implies that an intrinsic mechanism or a side-jump scattering mechanism of the SHE dominates the SOT generation [2,3]. However, in our case, the SOT enhancement is attributed to the OHE and/or OREE from the Co/CuN$_x$ bilayer. Though no related experimental study has been published before, a previous theoretical calculation study suggests that the OHC should be independent of the resistivity [15], which is consistent with what we observed here.

Further comparisons to other materials systems are shown in Fig. S4(b). We summarize the SOT conductivities of Pt/Co/MgO (control sample), W/CFB/MgO [50], and Ta/CFB/MgO [42] systems. It shows that the Pt/Co/CuN$_x$ structure, with the aid of the orbital current contribution, has the highest SOT conductivity.

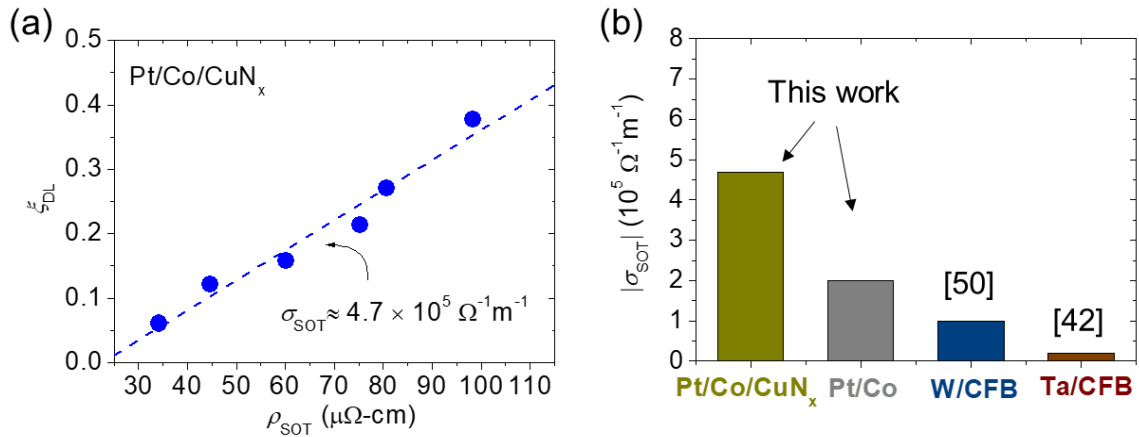

FIG. S4. (a) SOT efficiency as a function of resistivity of SOT layer. (b) SOT conductivity of Pt/Co/CuN$_x$, Pt/Co/MgO, W/CFB/MgO, and Ta/CFB/MgO systems.

## Supplementary Note 5. Intrinsic spin-orbit torque from the Co layer

According to the proposed scenario, the orbital angular momentum generated by the CuN$_{x_x}$ layer is first converted to the spin angular momentum through the spin-orbit coupling (SOC) of Co and then exerts a spin torque on the Co layer, whose sign is determined by the *L-S* conversion coefficient ($\eta$) of Co and is correlated to its own spin Hall ratio (SOT efficiency) [20]. Therefore, to examine the sign of $\eta$ in Co, we perform hysteresis loop shift measurement on a Co(2)/Pt(2)/Co(0.5)/Pt(2) magnetic heterostructure with an in-plane magnetized Co(2) layer as the spin Hall source (SOT layer) and a perpendicular magnetized Co(0.5) as the free layer, as shown in Fig. S5(a). The out-of-plane hysteresis loop shown in Fig. S5(b) confirms the PMA. Figure S5(c) shows the shifted hysteresis loops with $I_{dc} = \pm 2.5$ mA under $H_x = 800$ Oe, and the current-induced effective fields are summarized in Fig. S5(d). It is noted the negative slope ($H_z^{eff} / I_{dc}$) under $H_x = 800$ Oe indicates a SOT with negative sign generated from the Co layer. Therefore, the spin Hall ratio of our sputter-deposited thin film Co is determined to be negative, which corresponds to a negative $\eta$ of Co. Also noted that the Pt/Co/Pt trilayer has a minimal effect on this current-induced SOT effective field estimation [12].

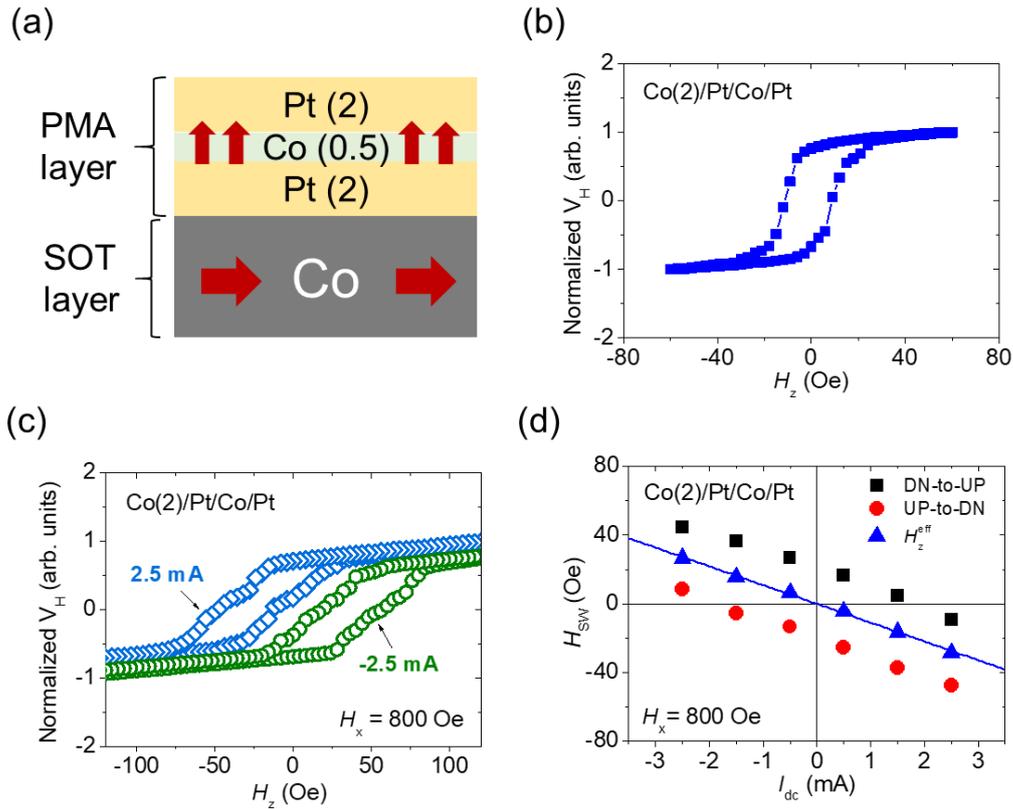

FIG. S5. SOT characterization and spin Hall ratio sign determination from a Co(2)/Pt(2)/Co(0.5)/Pt(2) device. (a) The illustration of Co(2)/Pt(2)/Co(0.5)/Pt(2) magnetic

heterostructure. (b) Out-of-plane hysteresis loop. (c) Shifted hysteresis loops under $I_{dc} = \pm2.5$ mA and $H_x$ = 800 Oe. (d) Current induced effective field under $H_x$ = 800 Oe.

## Supplementary Note 6. Orbital currents from the CuN$_x$ layer

We further prepare three series of samples to verify the orbital currents originated from the CuN$_x$ layer, namely: Pt($t$)/Co(1)/MgO(2), Pt($t$)/Co(1)/CuN$_x$(7), and CuN$_x$(7)/Pt($t$)/Co(1)/MgO(2) with $t$ ranges from 2 nm to 5 nm and Q of the CuN$_x$ layer is fixed at 25 %. First, Fig. S6(a) compares the Pt($t$)/Co/CuN$_x$ results with the Pt($t$)/Co/MgO results. The overall SOT efficiencies from the samples with CuN$_x$ caps are all greater than that from the standard Pt/Co/MgO samples, which confirms the existence of an OAM-SOT from the upper Co/CuN$_x$ structure with various Pt thicknesses, which corresponds to $\eta_{Co}\sigma_{OH}^{CuN_x} < 0$. And since $\eta_{Co} < 0$, the CuN$_x$ orbital Hall conductivity $\sigma_{OH}^{CuN_x} > 0$.

Next, the CuN$_x$ layer is placed underneath the Pt/Co/MgO structure to examine the orbital Hall scenario similar to a recent work [8], using Pt as the orbital-to-spin conversion layer with $\eta_{Pt} > 0$. The summarized SOT efficiency shown in Fig. S6 (b) shows that a strong enhancement is obtained at Pt = 2 nm and then gradually decreases with increasing Pt thickness. The result suggests that the CuN$_x$/Pt produces a positive-sign torque due to the orbital-to-spin conversion in the thin Pt layer. As a result, the overall SOT efficiency is enhanced by the positive product of $\eta_{Pt}\sigma_{OH}^{CuN_x} > 0$. As increasing the Pt layer thickness, the overall SOT efficiency decays due to (1) the short spin diffusion length in Pt and (2) the spin Hall effect (SHE) of Pt becomes the dominating SOT source. From these two sets of comparisons, we confirm that sizable orbital currents/torques can be efficiently generated from the CuN$_x$ layer with a positive OHC ($\sigma_{OH}^{CuN_x} > 0$) in adjacent to either the Co FM layer ($\eta_{Co} < 0$) or the Pt spacer ($\eta_{Pt} > 0$).

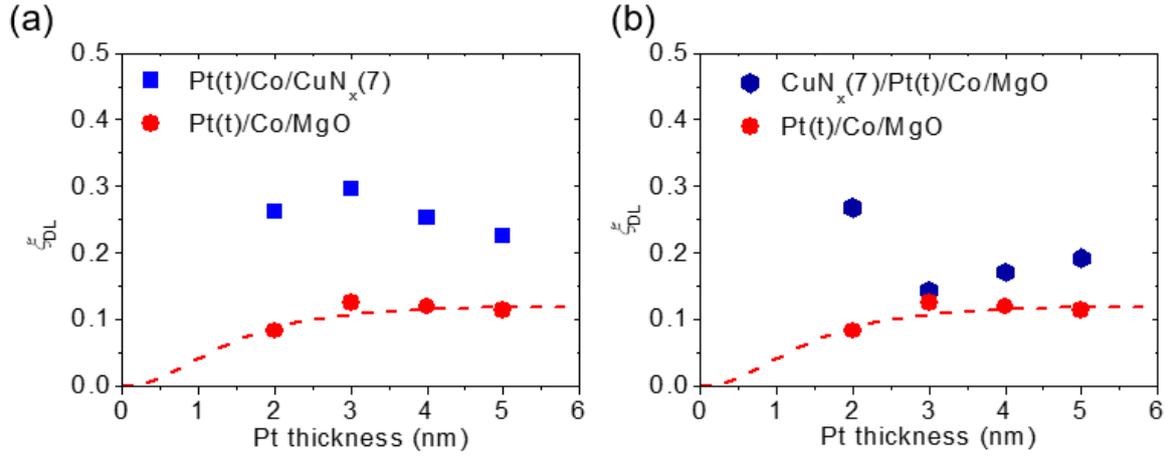

FIG. S6. (a) SOT efficiencies from Pt/Co/CuN$_x$ and Pt/Co/MgO samples. (b) SOT efficiencies from CuN$_x$/Pt/Co/MgO and Pt/Co/MgO samples. Dash line represents the fitting of spin diffusion model.

## Supplementary Note 7. Neuromorphic switching from the control sample

Here we show the similar neuromorphic/memristive switching behaviors as obtained from the Pt(3)/Co(1)/MgO(2)/Ta(2) control sample. The representative switching loops of using $I_{max}$ = 10, 12, 14, 16, 18, 20 mA are shown in Fig. S7(a). The critical switching current density is estimated to be ~ $18.5 \times 10^{10}$ A/m$^2$ from the $I_{max}$ = 20 mA switching loop. The switching percentage as a function of the maximum applied current density $J_{max}$ is summarized in Fig. S7(b) and the memristive window $\Delta J_{max}^{window}$ is found to be ~ $94 \times 10^{10}$ A/m$^2$.

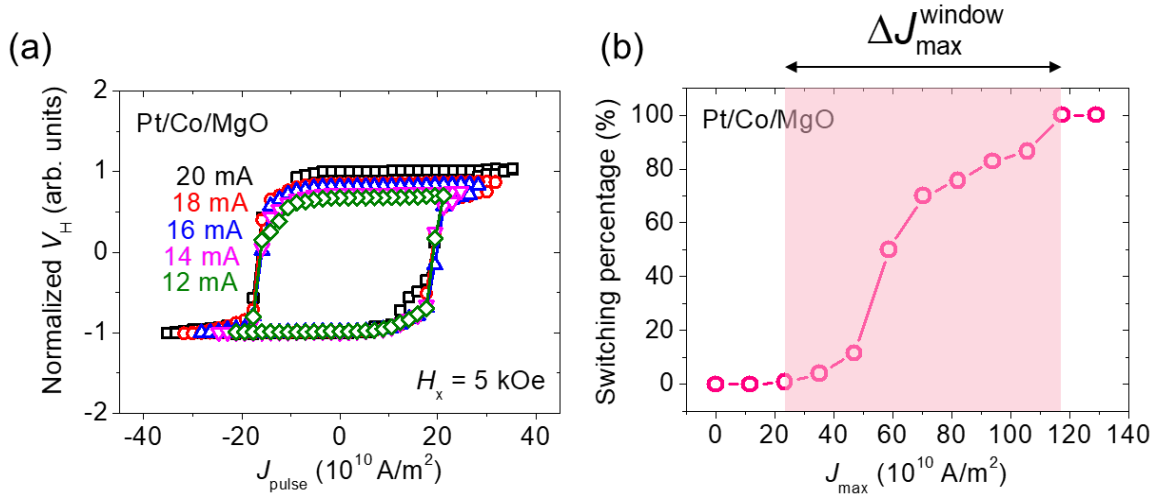

FIG. S7. Neuromorphic switching from a Pt(3)/Co(1)/MgO(2)/Ta(2) device. (a) Representative switching loops with various $I_{max}$. (b) Switching percentage with respect to the maximum applied current density. The memristive switching window is denoted as $\Delta J_{max}^{window}$.

## References

[44] H. K. Gweon, S. J. Yun, and S. H. Lim, A very large perpendicular magnetic anisotropy in Pt/Co/MgO trilayers fabricated by controlling the MgO sputtering power and its thickness, Sci. Rep. **8**, 1266 (2018).

[45] S. Bandiera, R. C. Sousa, B. Rodmacq, and B. Dieny, Asymmetric Interfacial Perpendicular Magnetic Anisotropy in Pt/Co/Pt Trilayers, IEEE Magn. Lett. **2**, 3000504 (2011).

[2] R. Karplus and J. M. Luttinger, Hall Effect in Ferromagnetics, Phys. Rev. **95**, 1154 (1954).

[3] M. I. Dyakonov and V. I. Perel, Current-induced spin orientation of electrons in semiconductors, Phys. Lett. A **35**, 459 (1971).

[15] T. Tanaka, H. Kontani, M. Naito, T. Naito, D. S. Hirashima, K. Yamada, and J. Inoue, Intrinsic spin Hall effect and orbital Hall effect in 4d and 5d transition metals, Phys. Rev. B **77**, 165117 (2008).

[50] W.-B. Liao, T.-Y. Chen, Y. Ferrante, S. S. P. Parkin, and C.-F. Pai, Current-Induced Magnetization Switching by the High Spin Hall Conductivity α-W, Phys. Status Solidi RRL **13**, 1900408 (2019).

[42] T.-Y. Chen, C.-T. Wu, H.-W. Yen, and C.-F. Pai, Tunable spin-orbit torque in Cu-Ta binary alloy heterostructures, Phys. Rev. B **96**, 104434 (2017).

[20] S. Lee *et al.*, Efficient conversion of orbital Hall current to spin current for spin-orbit torque switching, Commun. Phys. **4**, 234 (2021).

[12] T. Y. Chen, C. W. Peng, T. Y. Tsai, W. B. Liao, C. T. Wu, H. W. Yen, and C. F. Pai, Efficient Spin-Orbit Torque Switching with Nonepitaxial Chalcogenide Heterostructures, ACS Appl. Mater. Interfaces **12**, 7788 (2020).